# Design of an FPGA-based USB 3.0 device controller

**Zhe Ning**[a,1] **,Yunhua Sun**[a]

[a] *Institute of High Energy Physics, Chinese Academy of Sciences*
   *Beijing 100049, China*
   *E-mail*: ningzhe@ihep.ac.cn

ABSTRACT: The traditional USB 3.0 communication based on FPGA uses an external chip as a USB PHY or a USB controller including a USB PHY. This paper realizes a USB 3.0 controller using FPGA resources, in which FPGA logic realizes a serial interface engine, and an FPGA internal transceiver is a USB PHY. Used slices percent after implementation is 4.59% in Kintex-7 325t. The test result shows that the speed of USB 3.0 is more than 320 MB/s bulk-in and bulk-out transfers.

KEYWORDS: USB 3.0; FPGA; Transceivers.

# 1. Introduction

Because high precision and high-speed components are used in instruments based on FPGA, the requirement of high-speed transmission is necessary. The standard transmission scheme includes 125 MB/s Ethernet and USB 3.0. A widely used example of 125 MB/s Ethernet is SiTCP [1] which is realized based on a hardware protocol stack plus an external PHY. But USB 3.0 maximum speed is 625 MB/s, which is faster than 125 MB/s of Ethernet. Now the readout electronics integrated with Photomultiplier tubes (PMT) [2,3] is more and more popular. Still, the area of printed circuit boards (PCB) is limited by the size of PMT and power dissipation. At the same time, if dark rates of PMT which is nearly 50 kHz [4], needed to be studied deeply, the transmission bandwidth is very high. Because the window size of waveform sampling is 1000 points generally, the total size per second is 1000 points * 2 bytes/point * 50k = 100 MB/s which is very close to the limit of 125 MB/s of Ethernet. USB transmission schemes look better. But most USB schemes are realized using an external USB chip.

As shown in Figure 1, there are several typical schemes for USB in FPGA boards. In Architecture A, PHY, a serial interface engine and drivers are packaged into one chip, such as Cypress CY3014 [5] and FIDI FT600 [6], so that user logic in FPGA can send and receive data from the chip quickly. These chips are also called USB-FIFO chips. But they need scalability and cannot be a USB host. Architecture B, whose external chip includes a PHY, has more scalability and is more complicated because the other layers, such as the serial interface engine (SIE), are realized by FPGA logic. The most common external chip in architecture B is TUSB1310A [7] from Texas Instruments. It is unlucky that this chip has stopped production, and there is no replacement. This paper discusses Architecture C, which is the upgrade of Architecture B and tries to use an FPGA internal transceiver as a USB PHY to replace TUSB1310A. A smaller area of PCB for readout electronics could be adapted with PMT well.

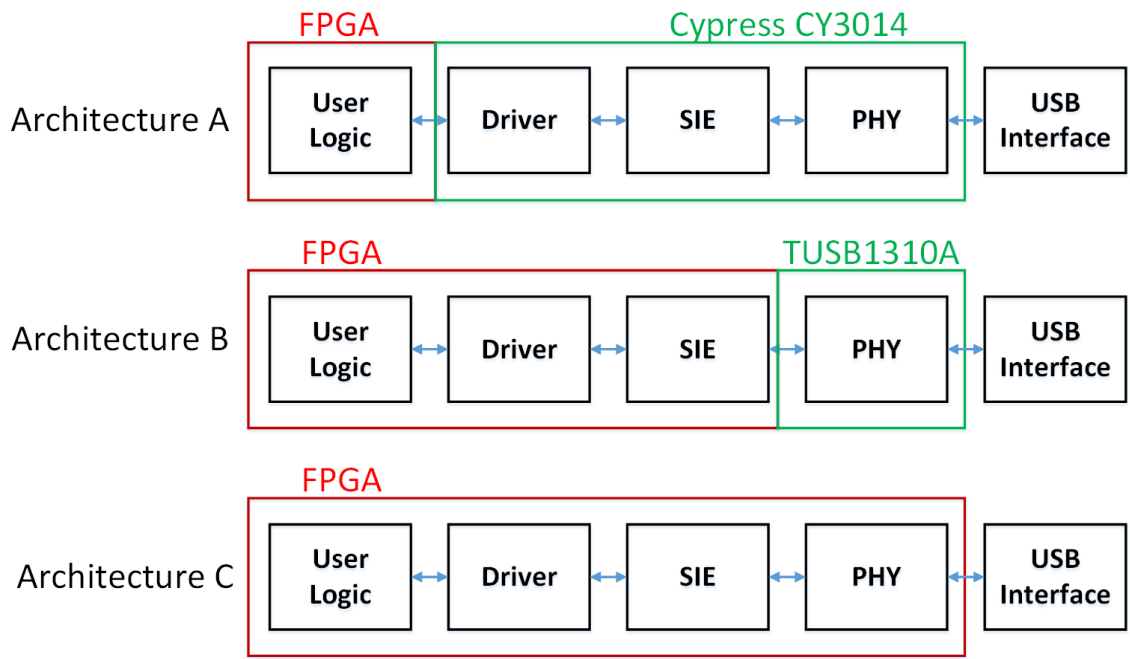

Figure 1 USB schemes for FPGA

## 2. System design

According to the document and support from Xilinx, Xilinx FPGA external transceivers are not used as a USB 3.0 PHY [8]. The primary function of USB PHY should be realized respectively. The standard USB 3.0 PHY includes a module transmitting Low-Frequency Periodic Signaling (LFPS) in which gigabit transceivers are turned off. The other is an SIE module communicating regular USB 3.0 data in which gigabit transceivers are turned on.

### 2.1 PHY modules

#### 2.1.1 LFPS definition

USB 3.0 PHY is similar to the PHY of PCIe and SATA, which should transmit Low-Frequency Periodic Signaling (LFPS) to initiate links or wake up the link partner in a low-power link state.

As Figure 2 shown, there are three parameters for LFPS: tPeriod, tBurst, and tRepeat. Because LFPS is a square wave, tPeriod is the period of a square wave whose minimum value is 20 ns, and the maximum value is 100 ns; tBurst is a period occupied by the transmission of continuous LFPS signal, and its value is dependent on LFPS types shown by Table 1 [9]; tRepeat are composed of tPeriod and the time of electric idle states by keeping two transmission differential wires at the same voltage, and its value is also depended on LFPS types. The most common LFPS type is polling, whose normal tBurst is 1.0 μs, and normal tRepeat is 10.0 μs.

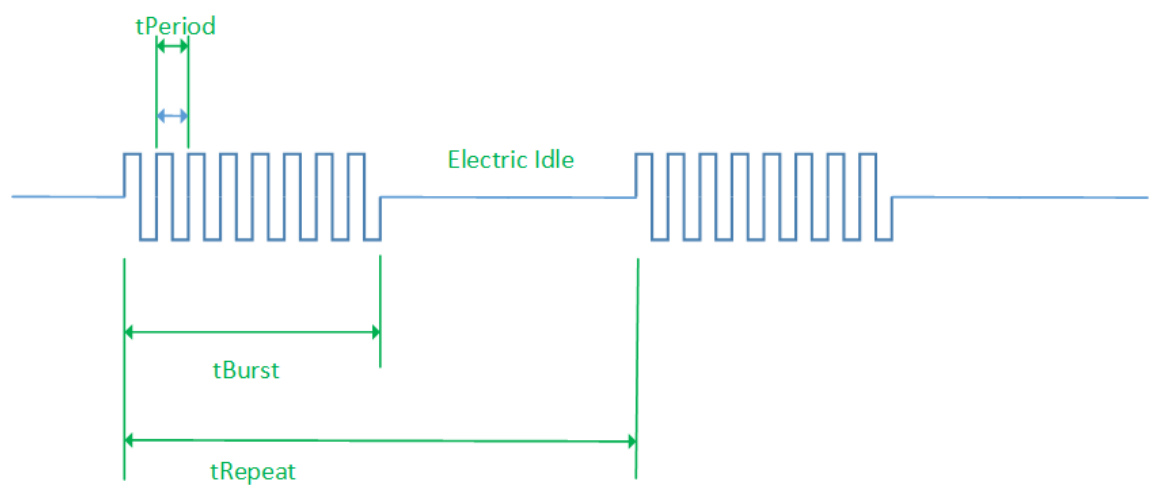

Figure 2 LFPS signaling

Table 1 LFPS Transmitter Timing for SuperSpeed Design

| | tBurst | | | | tRepeat | | |
|---|---|---|---|---|---|---|---|
| | Minimum | Normal | Maximum | Minimum Number of LFPS Cycles | Minimum | Normal | Maximum |
| Polling | 0.6 μs | 1.0 μs | 1.4 μs | | 6 μs | 10 μs | 14 μs |
| Ping | 40 ns | | 200 ns | 2 | 160 ms | 200 ms | 240 ms |

| tReset | 80 ms | 100 ms | 120 ms | | | | |
|---|---|---|---|---|---|---|---|
| U1 Exit | 600 ns | | 2 ms | | | | |
| U2 Exit | 80 μs | | 2 ms | | | | |
| U1 Wakeup | 80 μs | | 10 ms | | | | |

**2.1.2 Transmitting and receiving LFPS**

As discussed, some cases, such as link initiations or waking up link partners, should transmit LFPS by configuring some ports of transmitters shown in Figure 3. When initiating links, the generation conditions of LFPS are txpd = 2'b0 && rxpd = 2'b0 && TXDETECTRX = 1 && TXELECIDLE = 1 [10]; When waking up link partners, the generation conditions of LFPS are txppd = 2'b1 && rxpd = 2'b1 && TXELECIDLE = 0. For the detection of LFPS receiving, RXELECIDLE = 0 means LFPS detection, and RXELECIDLE = 1 means no LFPS detection [11].

  First, a square wave should be generated repeatedly during the tPeriod stages by toggling the wires between a differential '1' and '0' at a frequency. The data width of transceivers is 40 bits, in which 8b10b encoding is enabled but bypassed where TX8B10BEN = 1'b1 and TX8B10BBYPASS = 8'h0F. Due to the 625 MB/s speed of transceivers, the time occupied by each bit is 0.2 ns, so when tx_data = 40'hFF_FFFF_FFFF, the time of high level for one cycle is $0.2 \times 40 = 8$ ns, which is less than ten ns which is the half of one LFPS cycle, so the time of high level should be doubled at least. Figure 4 from oscilloscopes shows the tPeriod is 32 ns after doubling the time of the high level of square waves. If polling LFPS will be transmitted, tBurst consists of 32 tPeriod. Second, after the completion of tBurst, wires will come into an electric idle stage by asserting TXELECIDLE high.

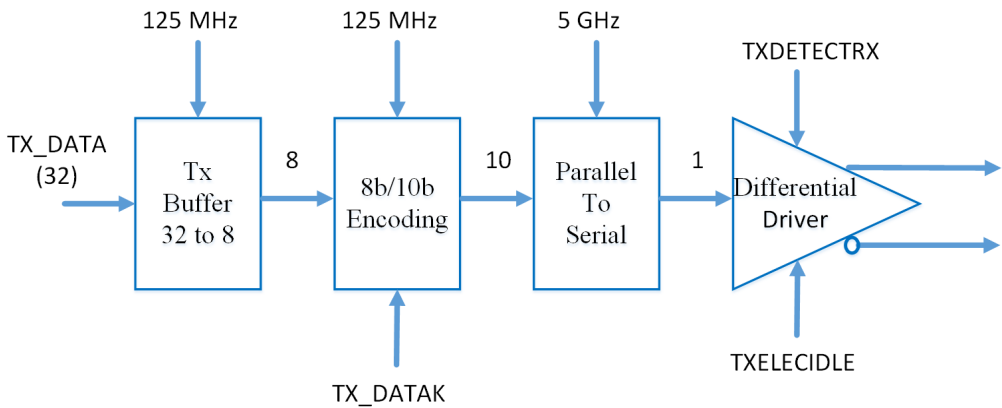

Figure 3 A transmitter structure

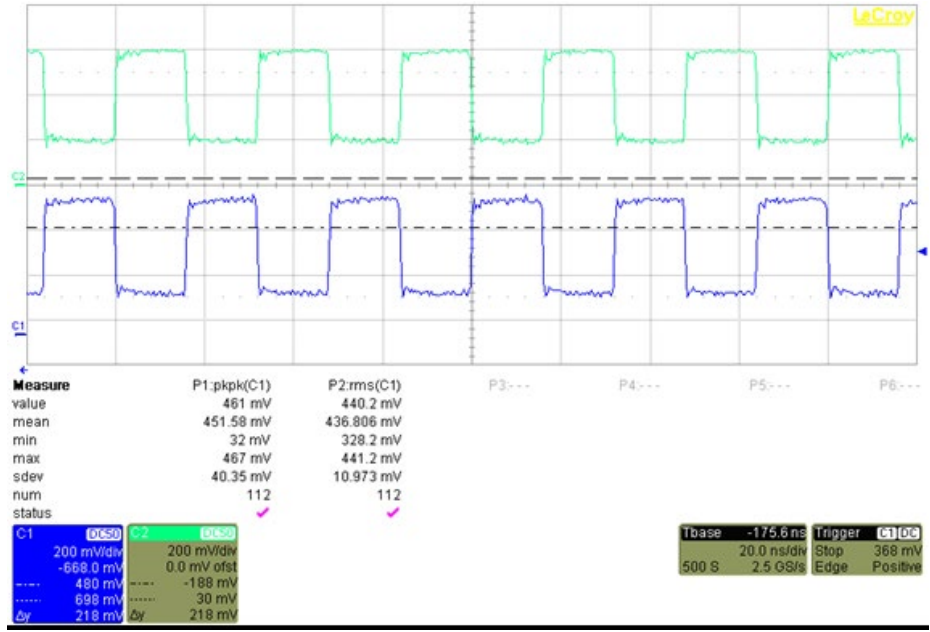

Figure 4 The tPeriod of LFPS signal captured by an oscilloscope

**2.2 SIE design**

As Figure 5 shown, SIE comprises three modules: the PIPE interface module, the link module, and the protocol module. The PIPE interface module is used for scrambling and descrambling data; The link module includes two parts: one is responsible for a link module and management, which is used to generate link command management packets, such as LGOOD and LCRD. The other one is Link Training Status State Machine (LTSSM), which is responsible for the transfer management from one state to another state, such as U0 and U1; the protocol module is used to generate tokens or data packets for the response of commands from a USB host when as a device or generate token or data packets to a USB device when as a host. For a USB device, an endpoints management module is needed to manage Endpoints 0, which is used for enumeration, Endpoint 1, and Endpoint 2, which is used for bulk-in and out. For a USB host, an enumeration management module is necessary to generate commands for enumerations after a device is inserted. A device driver module is also needed depending on the type of devices attached to hosts. Resource utilization is shown in Figure 6, and the percent of used slices in Kintex-7 325t is 4.59%.

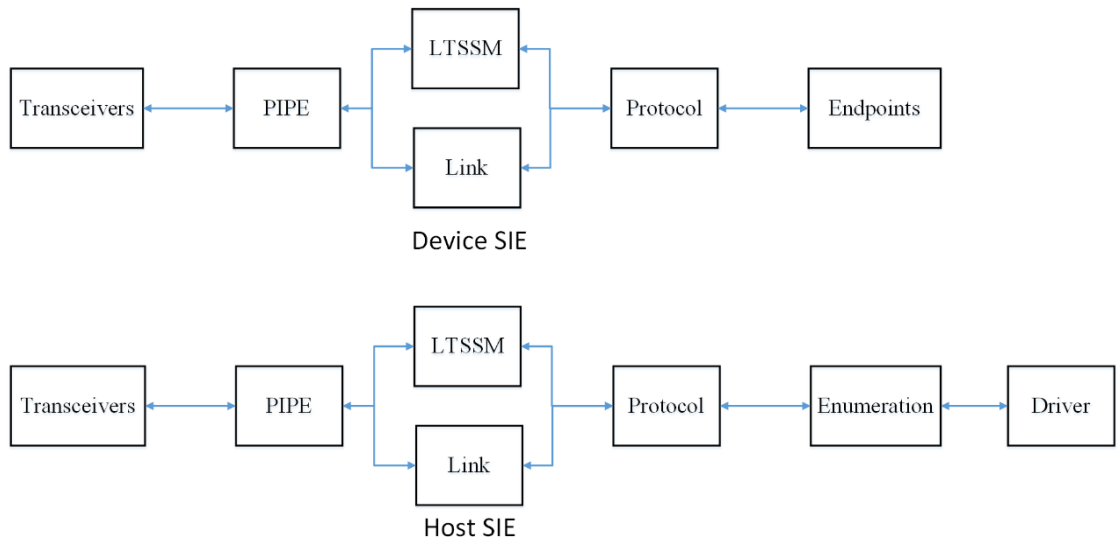

Figure 5 SIE design

| Slice LUTs (203800) | Slice Registers (407600) | F7 Muxes (101900) | Slice (50950) | LUT as Logic (203800) | LUT as Memory (64000) | Block RAM Tile (445) | DSPs (840) | Bonded IOB (500) | Bonded IPADs (50) | Bonded OPADs (32) | IBUFDS (480) | GTXE2_CHANNEL (16) |
|---|---|---|---|---|---|---|---|---|---|---|---|---|
| 5795 | 6340 | 6 | 2338 | 5665 | 130 | 9.5 | 1 | 4 | 4 | 2 | 1 | 1 |

Figure 6 Resource utilization of the USB controller

## 2.3 Initiation process

As shown in Figure 7, in the Rx.Detect state, PHY should be told to begin a receiver detection operation by asserting TXDETECTRX if the Signal phystatus is low. Signal phystatus is asserted High during receiver detection to indicate receiver detection completion [12]. So the state goes into a polling state. Transceivers should continue to transmit LFPS signals; On the other side, transceivers should check if the received LFPS signal is a polling signal. If it is, the transceivers are turned on and then transmit a TSEQ ordered set to the link partner for training the equalizer for a certain period. Following this, TS1 and TS2 ordered sets are sent as a handshake to finalize the link training and request info from the typical setup, such as a speedy response. The state will go into a U0 state from the polling state when all these are done. In the U0 state, at first, the USB controller will have a response for the enumeration request from a host and then come into bulk-in or bulk-out status.

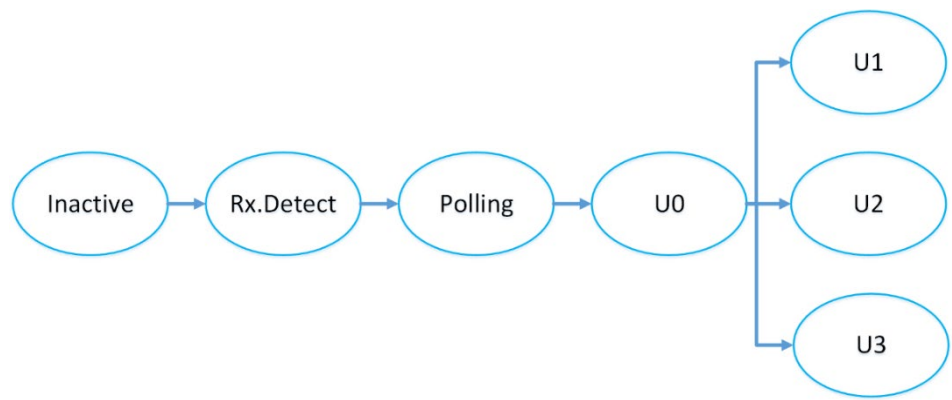

Figure 7 LTSSM High-Level States

## 3. Experiment results

### 3.1 Hardware setup

As shown in Figure 8, a KC705 board [13] is used for verification, and a mezzanine card named HiTech USB [14] is used for adding a USB interface connecting with GTX transceivers of Kintex-7 directly. More details about this connection are shown in Figure 9. It is noticeable that there should be a 0.1 uF capacitor in an RX termination, which is very common in a TX termination. The clock of the USB controller is from the output of the phase lock loop (PLL), whose input is associated with the TXOUTCLK port from transceivers whose reference clock is from a 200 MHz differential oscillator. So a FIFO is necessary between user logic and USB controllers due to different clock domains.

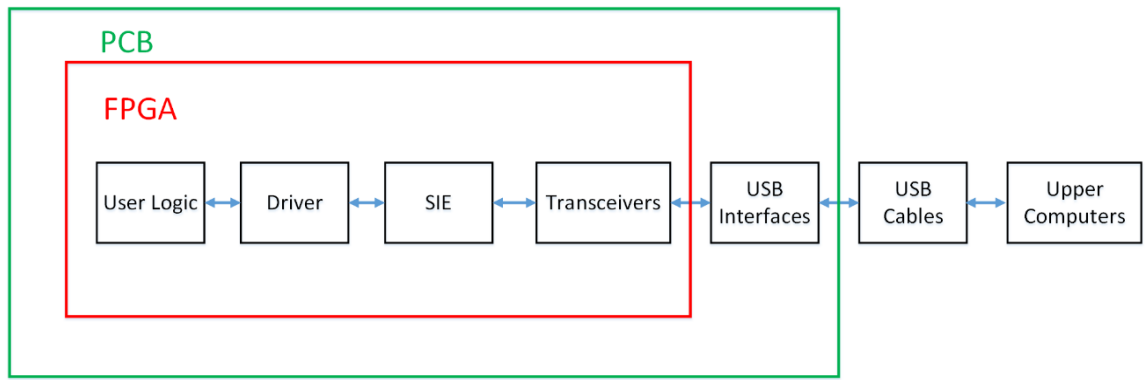

Figure 8 A hardware setup

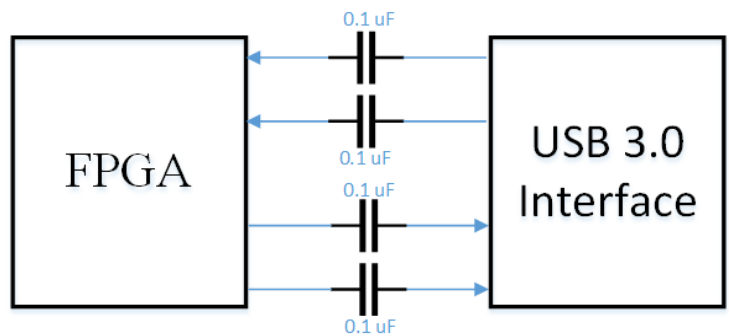

Figure 9 The FPGA USB 3.0 connector schematics

### 3.2 Speed tests

The speed tests include bulk-in and bulk-out transfer tests, which could be based on the upper computer software such as the cypress stream program or the third-party software of USB analyzers such as LeCroy adviser T3 [15]. For bulk-in tests shown in Figure 10 and Figure 11, both of cypress stream program and LeCroy adviser T3 show that the USB speed is more than 320 MB/s. At the same time, the USB analyzer also shows no bit error after transferring more than 3 TB of data, and then an estimated bit error rate is less than $10^{-13}$, which is less than $10^{-12}$ of the USB specification requirement. The bulk-out tests shown in Figure 12 and Figure 13 also indicate a similar result.

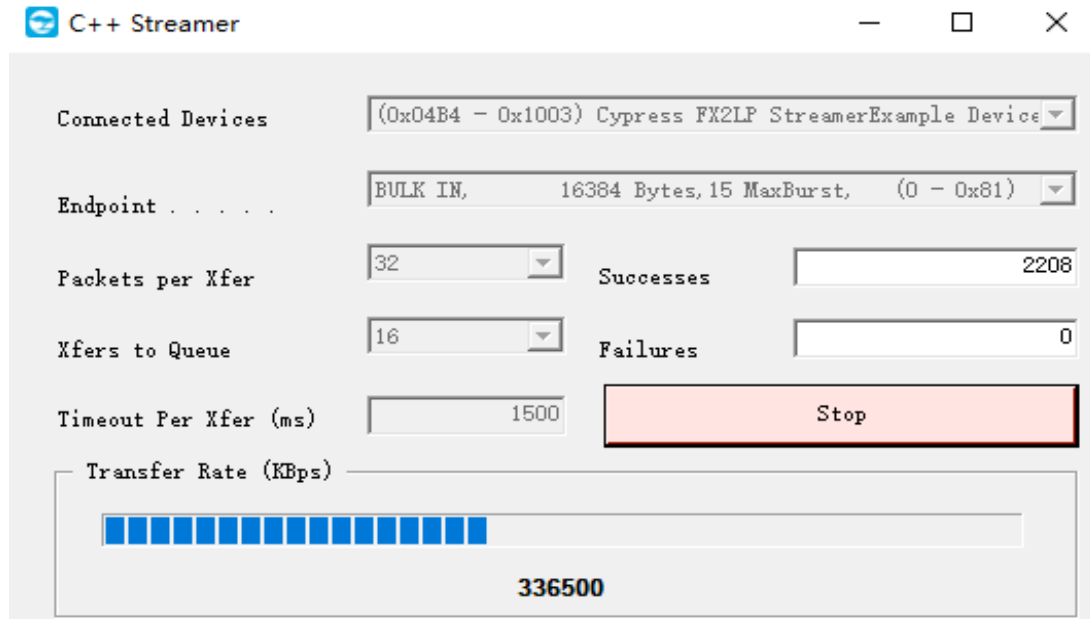

Figure 10 Bulk-in transfer test based on Cypress stream program

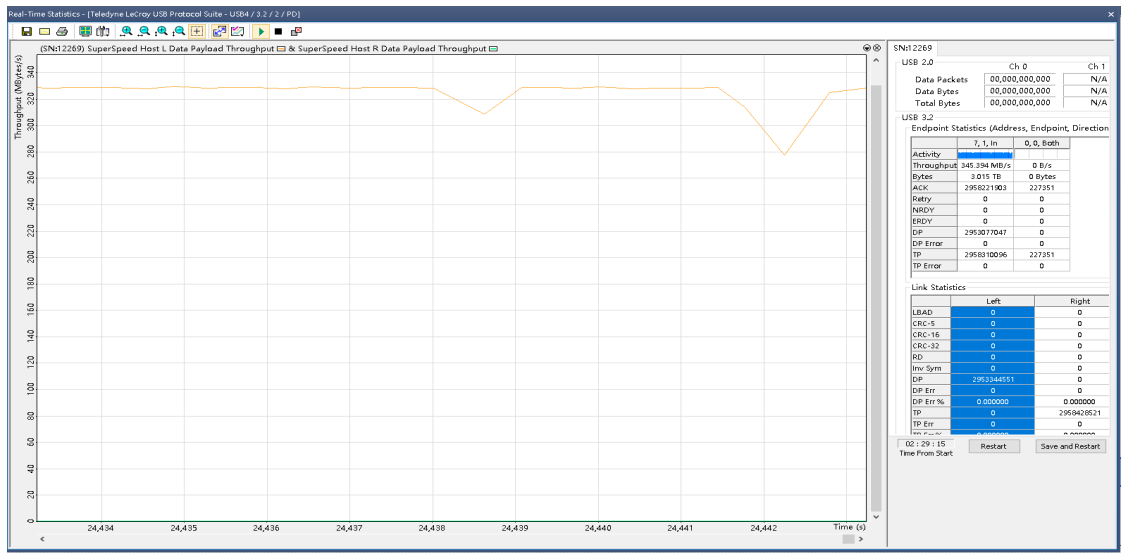

Figure 11 Bulk-in transfer test based on LeCroy adviser T3

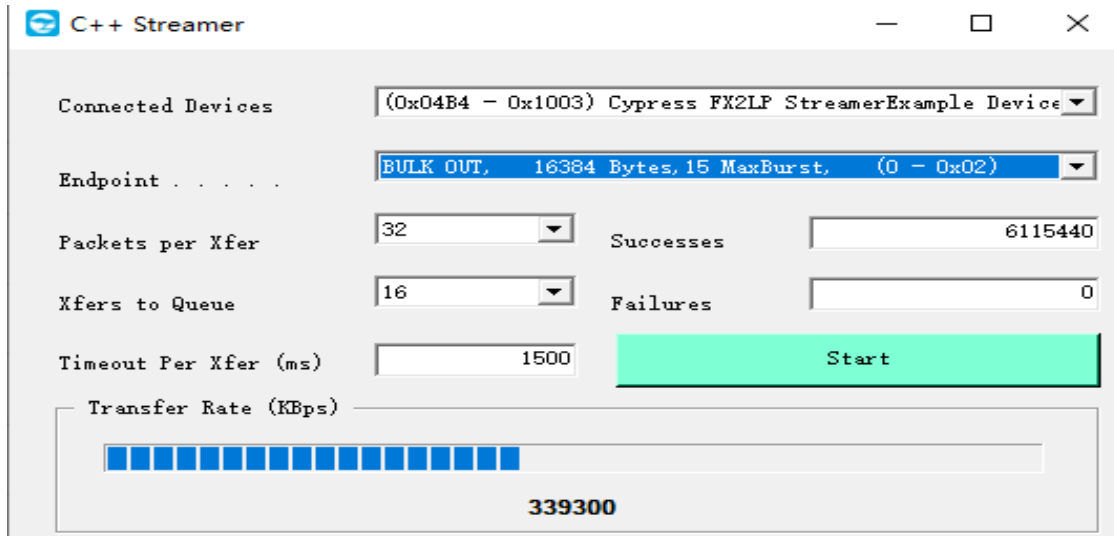

Figure 12 Bulk-out transfer test based on Cypress stream program.

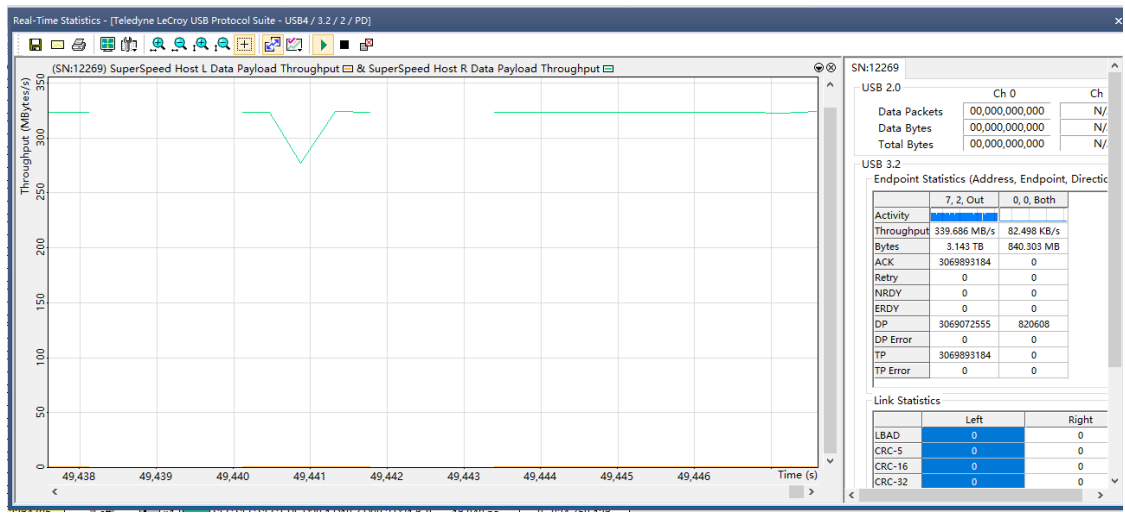

Figure 13 Bulk-out transfer test based on LeCroy adviser T3.

## 4. Summary

This paper proves that it is possible for USB 3.0 controllers, including PHY and SIE, based on FPGA internal transceivers and FPGA logic, and the test result shows the speed of bulk-in and bulk-out is more than 320 MB/s, and the bit error rate is less than $10^{-13}$. It is noticeable that the solution provided by this paper can be well dealt with USB 3.0 communication with upper computers. Still, if USB 2.0 compatibility is needed, an external USB 2.0 chip has to be added to the circuit.

## Acknowledgments

This project is supported by Beijing Natural Science Foundation (Grant No. 1214029).